\newfam\msbfam
\font\twlmsb=msbm10 at 12pt
\font\eightmsb=msbm10 at 8pt
\font\sixmsb=msbm10 at 6pt
\textfont\msbfam=\twlmsb
\scriptfont\msbfam=\eightmsb
\scriptscriptfont\msbfam=\sixmsb
\def\cj{\fam\msbfam}

\def\C{{\cj C}}

\def\Q{{\cj Q}}
\def\R{{\cj R}}

\def\Z{{\cj Z}}

\centerline{\bf TOPOLOGY AND COLLAPSE}

\

\centerline{\bf M. Socolovsky$^{(*)}$}  

\

\centerline{\it Instituto de Ciencias Nucleares, Universidad Nacional Aut\'onoma de M\'exico}
\centerline{\it Circuito Exterior, Ciudad Universitaria, 04510, M\'exico D. F., M\'exico} 
\centerline{\it and}
\centerline{\it Instituto de Astronom\'\i a y F\'\i sica del Espacio, Universidad de Buenos Aires-CONICET}
\centerline{\it C1428ZAA, Bs. As., Argentina} 
\

{\it We show that the collapse of the wave function of an entangled state of two spin ${{1}\over{2}}$ particles or two photons in the singlet state can be  geometrically understood as a change of fibre bundles.}

\


\

1. More than four decades of theoretical and experimental results, from the original paper of Bell$^1$ and experiments of Aspect {\it et al} $^2$, to the more recent violation of an inequality by Leggett$^3$ and Gr\"oblacher {\it et al} $^4$, strongly support the idea that quantum mechanics is a complete, non local, causal and non separable theory. (For a review, see for instance ref. 5.) A characteristic fact of this picture is the collapse of the wave function, a clearly non local/extended phenomenon. A global geometric description of it is then demanding. A step in this direction was given by Ne'eman $^6$: using the global character of parallel transport due to a connection in a principal fibre bundle over space-time, he suggested to visualize the collapse as a sort of spontaneous symmetry breakdown process. However, no explicit example of such a connection was given.

\

2. An ideal mathematical experiment could be the following: imagine two distant observers $A$ and $B$ living on $\R^2$. The plane is contractible, so its fundamental group is zero. At a certain time $t$, observer $A$ makes a small hole in the plane (it is enough to eliminate a single point). The whole space changes abruptly to a non simply connected space and the fundamental group becomes isomorphic to the integers. The space in which the spacelike separated (with respect to $A$) observer $B$ lives then changes instantaneously.

\

3. In some sense quantum mechanics, in particular through the collapse process, makes the previous imaginary experiment feasible. In fact, consider an entangled pair of two spin ${{1}\over {2}}$ particles flying appart from each other in the singlet spin zero state. (A similar analysis can be done with photons.) In a $z$-basis, the normalized spin wave function is $$\psi^<={{1}\over {\sqrt{2}}}(\uparrow_1\otimes\downarrow_2-\downarrow_1\otimes\uparrow_2). \eqno{(1)}$$ Since each spin belongs to $\C^2$, the spin Hilbert space $\cal H^<$ is $\C^2\otimes\C^2\cong\C^4$, and by normalization, $\psi^< \in S^7$, the 7-sphere. An overall $U(1)$ phase multiplies $S^7$, leaving as the physical space of states the set of lines through the origin or projective space of $\C^4$, $P(\C^4)$, namely, the complex projective 3-space $\C P^3$ (``ray space''). The initial wave function then involves the 3rd complex Hopf bundle: $$\xi_h^3: \ \ U(1)\to S^7 \buildrel\ {h_3}\over\longrightarrow \C P^3, \eqno{(2)}$$ where $$h_3(\vec{z})=\vec{z}\C^*, \eqno{(3)}$$ with $\C ^*=\C \setminus \{\vec{0}\}$. At the moment $t_0$ in which a measurement of the spin projection of, say, particle 2, along the (arbitrary) direction $\hat{n}$ is performed, with the result, say, $S_{({\hat{n}})_2}=-{{1}\over{2}}$, the wave function collapses: $$\psi^< \to \psi^>=(\hat{n})_1\otimes (-\hat{n})_2. \eqno{(4)} $$ Since the spin 2 fixes the direction of the spin 1, {\it i.e.}, once the measurement is performed, one spin contains all the information concerning the other spin, then the effective wave function can be considered a unit vector in $\cal H ^>$=$\C^2$ {\it i.e.} an element of $S^3$, the 3-sphere. Again, $U(1)$ gives an arbitrary phase on $S^3$ and the effective physical space of states becomes the ray or projective space of $\C^2$, $P(\C^2)$ namely, the complex projective line $\C P^1$ (homeomorphic to the 2-sphere $S^2$). Then the measurement process led us to the 1st complex Hopf bundle: $$\xi_h^1: \ \ U(1)\to S^3\buildrel\ {h_1}\over\longrightarrow \C P^1, \eqno{(5)}$$ with $$h_1(\vec{w})=\vec{w}\C^*. \eqno{(6)}$$ We can summarize the above change of bundle structure in the following commutative diagram: $$\matrix{U(1) & & U(1)\cr \downarrow & & \downarrow \cr S^3 & \hookrightarrow & S^7 \cr h_1 \downarrow & &  \downarrow h_3 \cr \C P^1 & \hookrightarrow & \C P^3 \cr & \buildrel\ {collapse}\over \longleftarrow & \cr      } \ \ ,$$ where $\hookrightarrow$ denotes the inclusion. 

\

4. $\xi_h^1$ and $\xi_h^3$ belong to the infinite sequence of $U(1)$-bundles $\xi_h^n: \ U(1)\to S^{2n-1}\buildrel\ {h_{n-1}}\over \longrightarrow\C P^{n-1}$, $n=1,2,...$ (the complex Hopf bundles) $^7$, with $\xi_h^n \hookrightarrow\xi_h^{n+1}$ and infinite limit $\xi_h^\infty: \ U(1)\to S^\infty \buildrel \ {h_\infty}\over \longrightarrow \C P^\infty$. ($S^\infty$ and $\C P^\infty$ are respectively given by the union of all $S^n$ and $\C P^n$ for finite $n$.) $\C P^\infty$, which is the clasifying space of $U(1)$, is an Eilenberg-Mac Lane space and has its homotopy concentrated in dimension 2; then there is a bijection given by the 1st Chern class, $c_1$:${\cal P}_{U(1)}(X)\to H^2(X;\Z)$, where ${\cal P}_{U(1)}(X)$ is the set of isomorphism classes of $U(1)$-bundles over $X$ and $H^2(X;\Z)$ is the second cohomology group of $X$ with integer coefficients. In particular, since $$H^2(\C P^1;\Z)\cong H^2(\C P^3;\Z)\cong \Z$$ then $${\cal P}_{U(1)}(\C P^1)\cong {\cal P}_{U(1)}(\C P^3)\cong \Z.$$ So, $\xi_h^1$ and $\xi_h^3$ are respectively the ``intersections'' of the sequence of Hopf bundles $\xi_h^n$ mentioned above, with the infinite sequences $P_k^3$ and $P_k^7$, $k\in \Z$, of $U(1)$-bundles over $\C P^1$ and $\C P^3$, with $P_1^3\cong S^3$ and $P_1^7\cong S^7$.

\

5. Equivalently, we can describe the collapse in terms of the associated vector bundles $\bar{\xi}_h^1$ and $\bar{\xi}_h^3$ due to the natural action of $U(1)$ on the complex numbers, $U(1)\times \C \to \C$, $(e^{i\rho},z)\mapsto e^{i\rho}z$, $\rho \in [0,2\pi)$. For $l=2$ and $l=4$, $$\bar{\xi}_h^{l-1}: \ \matrix{\C \cr \vert \cr S^{2l-1}\times_{U(1)}\C \cr \bar{h}_{2l-1}\downarrow \cr \C P^{l-1} \cr}$$ with $S^{2l-1}\times_{U(1)}\C =\{[(\vec{z},w)]\}_{(\vec{z},w)\in S^{2l-1}\times \C}$, $[(\vec{z},w)]=\{(\vec{z}e^{i\rho},e^{-i\rho}w)\}_{\rho\in[0,2\pi)}$, and $\bar{h}_{l-1}([(\vec{z},w)])=h_{l-1}(\vec{z})$. These line bundles are canonically isomorphic to the {\it canonical complex line bundles} $\gamma^c_{l-1}: \ \C-E(\gamma^c_{l-1})\buildrel\ {\pi_{c,l-1}}\over \longrightarrow \C P^{l-1}$ given by $\C P^{l-1}\times \C ^l \supset E(\gamma_{l-1}^c)=\{(\vec{z}\C ^*,\vec{w})\}_{(\vec{z},\vec{w})\in \C ^{l*}\times\C ^l}$ and $\pi_{c,l-1}((\vec{z}\C ^*,\vec{w}))=\vec{z}\C ^*$. The isomorphism is $$\matrix{\C & & \C \cr \vert & & \vert \cr S^{2l-1}\times_{U(1)}\C & \buildrel \psi_{l-1}\over \longrightarrow & E(\gamma_{l-1}^c) \cr \bar{h}_{2l-1}\downarrow & & \downarrow \pi_{c,l-1} \cr \C P^{l-1} & = & \C P^{l-1} \cr}$$ with $\psi_{l-1}([(\vec{z},w)])=(\vec{z}\C ^*,w\vec{z})$ and inverse $\psi_{l-1}^{-1}((\vec{z}\C ^*,\vec{w}))=[({{\vec{z}}\over{\vert\vec{z}\vert}},\lambda)]$, with $\lambda{{\vec{z}}\over{\vert \vec{z}\vert}}=\vec{w}$. 

\

A point of $E(\gamma_{l-1}^c)$ consists of a line through the origin in $\C ^l$ together with a vector on this line. In terms of $E(\gamma_{l-1}^c)$ one has the diagram: $$\matrix{\C & & \C \cr \vert & & \vert \cr E(\gamma_1^c) & \hookrightarrow & E(\gamma_3^c) \cr \pi_{c,1} \downarrow & & \downarrow \pi_{c,3} \cr \C P^1 & \hookrightarrow & \C P^3 \cr & \buildrel\ {collapse}\over \longleftarrow & \cr} \ \ .$$

\

6. As we said before, the bundles $\xi_h^1$ and $\xi_h^3$ belong to an infinite sequence $\xi_h^{n-1}$ of $U(1)$-bundles $$\matrix{S^1  & \cr  \downarrow & \cr ...\hookrightarrow S^{2n-1} \hookrightarrow ... \cr  \downarrow h_{n-1} \cr ...\hookrightarrow \C P^{n-1} \hookrightarrow ... \cr}$$ for $n=1,2...$, with $S^{2n-1}\subset \C P^n$, which is a filtration of the classifying bundle of the circle $\xi_h^\infty\equiv\xi_{U(1)}: U(1)\to S^\infty \buildrel {h_{\infty}}\over \longrightarrow \C P^\infty$ with $S^\infty =\cup_{n=1}^{\infty}S^n$ and $\C P^\infty=\cup_{n=1}^{\infty}\C P^n$. Since $U(1)$ is a compact Lie group, $\xi_h^\infty$ has a universal connection $^8$ given by the family of connections $\{\omega_{n-1}\}_{n=1}^\infty$, where $\omega_{n-1}$ is the connection on $\xi ^{n-1}_h$ induced by the metric on $\C ^n$, $<\vec{z},\vec{z} \ ^\prime>=\Sigma_{i=1}^n \bar{z}_i z_i^\prime$. At $\vec{e}=\pmatrix{1 \cr 0 \cr \vdots \cr 0}\in S^{2n-1}$, $\omega_{n-1}$ is given by $T_{\vec{e}}S^{2n-1}=V_{\vec{e}}\oplus H_{\vec{e}}$ with $V_{\vec{e}}=\{\pmatrix{it \cr 0 \cr \vdots \cr 0}\}_{t\in \R}$: vertical space at $\vec{e}$, and $H_{\vec{e}}=\{\pmatrix{0 \cr z_2 \cr \vdots \cr z_n}\}_{z_k\in \C , \ k=2,...,n}$: horizontal space at $\vec{e}$. In particular:

i) at $\vec{e}=\pmatrix{1 \cr 0 \cr 0 \cr 0}\in S^7$, $\omega_3$ is given by $V_{\vec{e}}=\{\pmatrix{it \cr 0 \cr 0 \cr 0}\}_{t\in \R}$ and $H_{\vec{e}}=\{\pmatrix{0 \cr z_2 \cr z_3 \cr z_4}\}_{z_k\in \C , \ k=2,3,4}$;  

ii) at $\vec{e}=\pmatrix{1 \cr 0}\in S^3$, $\omega_1$ is given by $V_{\vec{e}}=\{\pmatrix{it \cr 0}\}_{t\in \R}$ and $H_{\vec{e}}=\{\pmatrix{0 \cr z}\}_{z\in \C}$. 

The collapse process involves the transition $\omega_3 \to \omega_1$. Unfortunately, we can not give yet a physical meaning (if any) to this transition; however, this is the point of contact with the work of Ne'eman in ref. 6, where the assumed connection responsible for the collapse, though unspecified, was supposed to exist in space-time, rather than in spin (or polarization) space. Nevertheless, if Ne'eman's idea is correct, the connection in question (in spin space) has to be a pull-back of the universal one.

\

\

(*) On sabbatical year at IAFE

\

{\bf Acknowledgement}

\

This work was partially supported by the project PAPIIT IN113607, DGAPA-UNAM, M\'exico. The author thanks Professors F. Cukierman, G. L. Naber, and R. S. Huerfano for some remarks. 

\

{\bf References}

\

1. J. S. Bell, On the Einstein-Rosen-Podolsky paradox, {\it Physics} {\bf I}, 195-200 (1964).

\

2. A. Aspect, J. Dalibart and G. Roger, Experimental test of Bell's inequalities using time-varying analyzers, {\it Physical Review Letters} {\bf 49}, 1804-1807 (1982). 

\

3. A. J. Leggett, Nonlocal Hidden-Variable Theories and Quantum Mechanics: An Incompatibility Theorem, {\it Foundations of Physics} {\bf 33}, 1469-1493 (2003).

\

4. S. Gr\"oblacher, T. Paterek, R. Kaltenbaek, C. Brukner, M. Zukowski, M. Aspelmeyer and A. Zeilinger, An experimental test of non-local realism, {\it Nature} {\bf 446}, 871-875 (2007).

\

5. A. Aspect, ``Bell theorem: the naive view of an experimentalist'', in {\it Quantum [Un]speakables-From Bell to Quantum information}, ed. by R. A. Bertlmann and A. Zeilinger, Springer (2002).

\

6. Y. Ne'eman, Classical geometric resolution of the Einstein-Podolsky-Rosen paradox, {\it Proceedings of the National Academy of Sciences, USA} {\bf 80}, 7051-7053 (1983); The Problems in Quantum Foundations in the Light of Gauge Theories, {\it Foundations of Physics} {\bf 16}, 361-377 (1986).

\

7. N. Steenrod, {\it The Topology of Fibre bundles}, Princeton University Press, Princeton (1951).

\

8. M. S. Narasimhan and S. Ramanan, Existence of universal connections, {\it American Journal of Mathematics} {\bf 83}, 563-572 (1961).

\

\

\

e-mails: socolovs@nucleares.unam.mx, socolovs@iafe.uba.ar

\end

\

\

\

\

\

\
As is well known, the following three hypotesis: i) discretization with integers (or ''quantization'') of electric charge, ii) quantum mechanics, and iii) gauge invariance, imply that the gauge group of quantum electrodynamics (QED) is $U(1)$ and not $\R$.

In fact, consider the pure electromagnetic 4-potential $A_\mu$ in Minkowski 4-dimensional spacetime $M^4$ (the manifold); the transformation $$A_\mu \to A_\mu+\partial_\mu \lambda \eqno{(1)}$$ with $\lambda:M^4\to \R$ a differentiable ($C^\infty$ function), leaves the Maxwell equations in the vacuum invariant. (It is enough to consider for $A_\mu$ the classical situation, its quantization does not modify any of the following results; also, we work in the natural system of units $\hbar=c=1$, then $[\lambda]=[lenght]^0$.) Thus, at the level of the pure electromagnetic field, the gauge group at each spacetime point is $\R$, and the gauge group in the manifold is ${\cal G}=\R^{M^4}=\{smooth \ functions \ M^4\to \R \}$. On the other hand, the Schroedinger equation of a non relativistic charged scalar (or pseudoscalar) particle coupled to $A_\mu$ (the complications of relativity and/or spin are here irrelevant) is invariant if simultaneously with (1) the wave function changes as $$\psi(x)\to \psi^\prime(x)=e^{iq\lambda(x)}\psi(x) \eqno{(2)}$$ where $q$ is the electric charge of the quantum particle. Let $$q=ne_0, \ n\in \Z \eqno{(3)}$$ where $e_0>0$ is a fundamental electric charge. Then, the transformation (2) is $$\psi^{\prime}(x)=e^{ine_0\lambda(x)}\psi(x)=e^{i2\pi n\Lambda_0(x)}\psi(x) \eqno{(4)}$$ where $$\Lambda_0(x)=\tilde{e}_0 \lambda(x) \eqno{(5)}$$ with $$\tilde{e}_0={{e_0}\over {2\pi}}. \eqno{(6)}$$ (In terms of $\tilde{e}_0$, the fine structure constant is given by $\alpha_0={{e_0^2}\over {4\pi}}=\pi \tilde{e}_0^2.$) 

Let $m$ be an integer and define the function $$\Lambda_m=\Lambda_0+m. \eqno{(7)}$$ In (4), call $\psi^\prime\equiv \psi_0^\prime$; then $$\psi_m^\prime(x)=e^{2\pi n \Lambda_m(x)}\psi(x)=e^{i2\pi n(\Lambda_0(x)+m)}\psi(x)=e^{i2\pi n \Lambda_0(x)}e^{i2\pi nm}\psi(x)=\psi_0^\prime(x) \eqno{(8)}$$ {\it i.e.} $$\Lambda_m \sim \Lambda_0 \ \ or \ \ \tilde{e}_0\lambda+m \sim \tilde{e}_0 \lambda, \eqno{(9)}$$ equivalently, at each point $x$, $$\tilde{e}_0\lambda(x)+m \sim \tilde{e}_0\lambda(x). \eqno{(10)}$$ Then, at each $x$, the symmetry group of electrodynamics is not $\R$ but its quotient by $\Z$: $$G_{QED}={{\R}\over {\Z}}\cong U(1). \eqno{(11)}$$ In $M^4$, ${\cal G}_{QED}\cong U(1)^{M^4}$. Topologically, $U(1)\cong S^1$, the 1-sphere. (Renormalization in quantum field theory merely changes $e_0$ to ${e_0} _{ren}$.) It is interesting to mention that in the transition $\R \to S^1 \cong \R \cup \{\infty\}$ there is not only a topology change $\tau_\R \to \tau_{S^1}=\tau_\R \cup \{V\cup \{\infty\}\}$, where $V$ is open in $\R$ and $\R\setminus V$ is compact in $\R$, but also a homotopy change: $\R$ is homotopically equivalent to a point while $S^1$ is not contractible; in particular, the fundamental group of the circle is $\Pi_1(S^1)\cong \Z$, with all other homotopy groups equal to zero. The {\it mathematical} appearance of the integers is a consequence of their {\it physical} presence in (3).

\

As is well known, the change $G_{QED}=\R \to G_{QED}=U(1)$ allows for the existence of non trivial $U(1)$-bundles over $S^2$, the so called {\it monopole bundles} $U(1)\to P^3_g \to S^2$ with $g \in \Z$, and therefore to monopole connections with magnetic charges given by the Chern numbers $g$ (Trautman, 1977; Choquet-Bruhat {\it et al}, 1982). The equivalence is mathematical, though not physical, since no Dirac monopole has yet been found in Nature. Also, the change $\R \to S^1$ remits to the classifying (or universal) bundle of the group $\Z$: $$\xi_\Z: \Z\to\R \buildrel{exp(2\pi i-)}\over \longrightarrow S^1 \eqno{(12)}$$ which plays a crucial r\^ole in the real time path integral formulation of quantum mechanics and quantum field theory. In particular, (12) implies that $\R$ is a non trivial extension of $\Z$ by $S^1$; the associated short exact sequence of groups and group homomorphisms $$0\to \Z \to \R \to S^1 \to 0 \eqno{(13)}$$ does {\it not} split, since a non trivial principal bundle has no (global) section. Another way to see this, is that if $\R$ were a semidirect sum $\Z \odot S^1$, then $\R$ would be topologically equivalent to $\Z \times S^1$, which is false. 

\

{2. \bf Fractional quark charges}

\

The derivation in section 1 was based on the assumption that electric charges are integer multiples of the unit charge $e_0$. Quarks, on the other hand, in the standard model, have electric charges $\pm{{2}\over {3}}e_0$ and $\pm{{1}\over {3}}e_0$, and couple electromagnetically to $A_\mu$ as the usual charged leptons (electrons, muons, taus and their antiparticles) do. This fact is independent of the quarks being confined. Then, a natural question to ask is if the usual $U(1)$ symmetry group of the electromagnetic interaction survives to those fractional charges and couplings. The idea is {\it to insist for the electromagnetic gauge group to be the quotient of $\R$ by one of its subgroups.} 

A first attempt in this direction is to consider the possibility of arbitrary fractional electric charges {\it i.e.} $q={{m}\over{n}}e_0$ with $m,n\in \Z$ and $n\neq 0$. But then the QED group would be ${{\R}\over {\Q}}$ which is isomorphic to $\R$ (Rose, 1978). (One can also show that ${{\R}\over {\Q}}\cong {{U(1) \over T(U(1))}}$ where, if $G$ is a group, then $T(G)$ is the torsion subgroup of $G$ (the elements of $G$ different from the identity which are of finite order).) But then $G_{QED}$ would be $\R$ and Dirac monopoles could not exist. 

Instead, the subgroup $G_Q$ of $\Q$ generated by $\{\pm{{2}\over {3}},\pm{{1}\over{3}}\}$, {\it i.e.} the smallest subgroup of $\Q$ that contains the quark charges, is $$G_Q=<\{\pm{{2}\over{3}},\pm{{1}\over{3}}\}>=\{{{m}\over{3}}\}_{m\in\Z} \eqno{(14)}$$ which is trivially isomorphic to $\Z$ through $${{m}\over{3}}\mapsto m. \eqno{(15)}$$ (Mathematically, it is enough to take $+{1 \over 3}$ or $-{1 \over 3}$ as generator of $G_Q$.) 

\

Then, for the electromagnetic symmetry group one recovers $U(1)$: $$G_{QED}\cong {{\R}\over{G_Q}}\cong{{\R}\over{\Z}}\cong U(1). \eqno{(16)}$$ This shows the compatibility between the known quark fractional charges, its generated group, and the symmetry group of the electromagnetic interaction. 

\

{\bf 3. Magnetic monopoles}

\

The Dirac quantization condition for an electric charge $q$ moving in the presence of a magnetic charge $g$ is given by (Dirac, 1931) $$qg=2\pi n, \ n\in \Z^*=\Z\setminus \{0\}, \eqno{(17)}$$ with $q={{m}\over{3}}e_0$ and  $m\in \Z$; for the magnetic charges it gives the values $$g_\rho=\rho g^\prime_0 \eqno{(18)}$$ with $\rho={n \over m} \in \Q ^*=\Q \setminus \{0\}$ and $g^\prime_0={3\over \tilde{e}_0}$. Then, all rational multiples of the fundamental magnetic charge $$g^\prime_0\equiv {6\pi \over e_0} \eqno{(19)}$$ are allowed. In the integer case, the same result holds with $g^\prime_0$ replaced by $g_0={1 \over \tilde{e}_0}$. 

\

{\bf 4. Dyons}

\

Assume now that there exist two dyons respectively carrying electric and magnetic charges $q_1, g_1$ and $q_2,g_2$ ($q_i,g_j \in \R$); they move with respect to each other following the Schroedinger equation. The Dirac quantization condition (17) is replaced by (Schwinger, 1969) $$q_1g_2-q_2g_1=2\pi n, \ n\in \Z. \eqno{(20)}$$ If $q_j={m_j \over 3}e_0$ with $m_j \in \Z, \ j=1,2$, then $m_1g_2-m_2g_1=ng^\prime_0$ and therefore, if $m_1 \neq 0$, $g_2={1 \over m_1}(m_2 g_1+ng_0^\prime)$ which admits rational and irrational solutions for both $g_1$ and $g_2$: $$if \ g_1=\rho_1 g^\prime_0 \ with \ \rho_1 \in \Q \ then \ g_2={1 \over m_1}(m_2 \rho_1 +n)g^\prime _0 \equiv \rho_2 g^\prime_0 \ with \ \rho_2 \in \Q, \eqno{(21)}$$ $$if \ g_1=\lambda g^\prime_0 \ with \ \lambda \in \R \ then \ g_2={1 \over m_1}(m_2 \lambda + n)g^\prime_0 \equiv rg^\prime_0 \ with \ r\in \R \setminus \Q. \eqno{(22)}$$ 

\ 

(If $r\in \Q$ then $\lambda={1 \over m_2}(rm_1+n)\in \Q$ if $m_2 \neq 0$ which is a contradiction.) If $m_2=0$ we reobtain (17) and $g_2$ is given by (18). Then, if dyons exist with their electric charges in $G_Q$, then their corresponding magnetic charges are real (rational or irrational) multiples of the fundamental magnetic charge $g^\prime_0$. As with monopoles, the same result holds in the integer case with $g^\prime_0$ replaced by $g_0$. 

\

{\bf Acknowledgement}

\

M. S. thanks the hospitality of the Departamento de F\'\i sica Te\'orica de la Universidad de Valencia, Spain, where part of this work was performed.

\

{\bf References}

\

Choquet-Bruhat, Y., De Witt-Morette, C. and Dillard-Bleick, M. (1982). {\it Analysis, Manifolds and Physics, Part I: Basics}, North Holland, Amsterdam: p. 408. 

\

Dirac, P. M. A. (1931). Quantized Singularities in the Electromagnetic Field, {\it Proceedings of the Royal Society of London A} {\bf 133}, 60-72.

\

Rose, J. S. (1978). {\it A Course on Group theory}, Dover, New York: p. 48. 

\

Schwinger, J. (1969). A Magnetic Model of Matter, {\it Science} {\bf 165}, 757-761.

\

Trautman, A. (1977). Solutions of the Maxwell and Yang-Mills Equations Associated with Hopf Fibrings, {\it International Journal of Theoretical Physics} {\bf 16}, 561-565.

\

\

\

\

\

\

e-mails:
alicia@nuclecu.unam.mx, socolovs@nuclecu.unam.mx

\end